\documentclass{webofc}

\usepackage[varg]{txfonts}   
\usepackage{hyperref}
\usepackage{url}
\hypersetup{colorlinks=true,citecolor=blue,urlcolor=blue,linkcolor=blue}
\begin{document}
\title{Recent advancements in the strongly coupled many-body theory 
for nuclear spectral computation
}
%
%

\author{\firstname{Elena} \lastname{Litvinova}\inst{1}\fnsep\thanks{\email{elena.litvinova@wmich.edu}} 
}
\institute{Western Michigan University, Kalamazoo, MI 49009 
          }

\abstract{Some recent advancements of the nuclear many-body theory and selected results on nuclear giant and pygmy resonances are presented. The theory is compactly reviewed, with a special focus on the emergent scale of the quasiparticle-vibration coupling (qPVC), which carries the order parameter associated with the qPVC vertex, and an efficient treatment of the nuclear many-body problem organized around the qPVC hierarchy.  
Self-consistent numerical solutions of the relativistic Bethe-Salpeter-Dyson equation for the nuclear response function in medium-heavy nuclei are discussed. The presented update on the pygmy dipole resonance focuses on establishing the formation of its two-component structure as a result of the fragmentation of the low-energy dipole mode due to the qPVC and its mixing with the similarly fragmented giant dipole resonance. 
The centroid of the isoscalar giant monopole resonance is also linked to qPVC effects, particularly to its sensitivity to the coupling of the collective breathing mode to the lowest quadrupole vibrations, which is enhanced by quadrupole collectivity. The resolution of the long-standing "fluffiness" puzzle regarding the compressibility of open-shell tin isotopes is summarized.
The recently developed thermal variant of the superfluid response theory is briefly introduced and continuously linked to the description of the isoscalar monopole response at finite temperature with the prospect of refining the temperature-dependent nuclear equation of state. 
%
}
\maketitle
\section{Introduction}
\label{intro}
Nuclear structure theory is a mature field still striving for spectroscopic accuracy of computation required for numerous applications from nuclear science and technologies to fundamental physics frontiers and new physics searches. The quest for a predictive theory for the low-energy nuclear structure phenomenology is still far from complete, and it is even unclear to what extent such a theory can reach the standards required by modern applications. The nuclear structure physics is sensitive to
several energy scales that can only be approximately decoupled within the effective theory frameworks: quantum hadrodynamics (QHD) is not continuously calculable from quantum chromodynamics, and the strongly-correlated nuclear medium is not satisfactorily calculable from the QHD without introducing new low-energy constants at each scale. The separation of scales thus comes at the price of uncontrolled errors, which prohibit achieving spectroscopic accuracy. 

The recent effort was therefore focused on the analysis and refined computation of the dynamical kernels of fermionic equations of motion (EOMs), which formally bridge the QHD and nuclear medium scales in terms of the time-dependent induced interaction.
The use of the vacuum nucleon-nucleon (NN) interactions \cite{PapakonstantinouRoth2009,Knapp:2014xja,Knapp:2015wpt} and higher configuration complexities were explored \cite{Ponomarev1999,LoIudice2012,Savran2011,Tsoneva:2018ven,Lenske:2019ubp} in the studies of the response of medium-heavy nuclei. The latter indicated that complex configurations are responsible for both gross and fine spectral details.  The "ab initio" EOMs, relying on only the bare NN interaction as an input, \cite{AdachiSchuck1989,DukelskyRoepkeSchuck1998,LitvinovaSchuck2019} combined with the effective meson-exchange NN interaction \cite{VretenarAfanasjevLalazissisEtAl2005,PaarNiksicVretenarEtAl2004a,PaarVretenarKhanEtAl2007}, constitute our relativistic nuclear field theory (RNFT) \cite{LitvinovaRingTselyaev2007,LitvinovaRingTselyaev2008,MarketinLitvinovaVretenarEtAl2012}. It introduced self-consistent computation of the response of medium-light nuclei, where the wave functions of the excited states contain correlated three-particle-three-hole ($3p3h$), or six-quasiparticle ($6q$), configurations \cite{LitvinovaSchuck2019,Litvinova2023a,Novak2024} organized by qPVC. 
On the formal level, this work connected the qPVC with the ab-initio dynamical kernels of the EOMs for the nucleonic correlation functions (CFs) in nuclear medium \cite{RingSchuck1980,Schuck1976,AdachiSchuck1989,Danielewicz1994,DukelskyRoepkeSchuck1998,SchuckTohyama2016,LitvinovaSchuck2019,LitvinovaSchuck2020}. We showed explicitly that {\it in the regime of the NN intermediate coupling, there emerges a new order parameter associated with the qPVC vertex, which takes over the power counting dominating the NN and few-nucleon systems} and thus can be used to produce a hierarchy of converging approximations to nucleonic CFs \cite{Novak2024}. The leading-order qPVC kernels can be mapped to the non-relativistic NFT and related to the quasiparticle-phonon model (QPM) kernels \cite{BohrMottelson1969,BohrMottelson1975,Broglia1976, BortignonBrogliaBesEtAl1977,BertschBortignonBroglia1983,Soloviev1992}, in the fully microscopic EOM framework \cite{LitvinovaSchuck2019,LitvinovaSchuck2020}. The latter opened the door to generalizing the NFTs to more complex configurations, although establishing a transparent link between the bare and effective NN interactions remains challenging.

Our Ref. \cite{Litvinova2022} was focused on the details of the superfluid extension of the two-fermion EOM approach by working out the complete formalism with both static and dynamical interaction kernels in the space of the Bogoliubov quasiparticles and keeping the 2$\times$2 matrix structure of the Hartree-(Fock)-Bogoliubov (H(F)B) basis. 
The major formal step was the unification of the particle-hole ($ph$) and particle-particle ($pp$) channels in one EOM with the qPVC dynamical kernel, unifying normal and pairing phonons and thus extending the applicability of the complete theory and qPVC approximations to open-shell nuclei, which constitute the major part of the nuclear landscape.
The recent numerical implementations of RNFT for the nuclear response with $4q$ ($2q\otimes phonon$) configurations addressed the nuclear compressibility puzzle \cite{Litvinova2023}, the fine structure of the giant monopole resonance \cite{Bahini2024}, giant dipole resonance systematics \cite{Markova2024}, and new aspects of splitting of the pygmy dipole resonance \cite{Markova2024} in the leading qPVC approximation. The next and so far maximally feasible complexity includes $6q$ configurations organized into $2q\otimes 2phonon$ structure. This approach demonstrated an improved performance as compared to the correlated $4q$ ($2q\otimes phonon$) one for spherical nuclei in the calcium, nickel, and tin mass regions \cite{LitvinovaSchuck2019,Litvinova2023,Novak2024} and, more importantly, for moderately deformed and transitional medium-mass species \cite{Muescher2024,Li2024}. The latter capability is the new feature of the RNFT, which enables adequate capturing of the deformation effects dynamically by high-complexity qPVC configurations, retaining the technical advantage of working in spherical bases. 
The first $ph\otimes phonon$ finite-temperature variants of the response theory beyond the "standard" quasiparticle random phase approximation (QRPA) \cite{Sommermann1983} were worked out for the $ph$ and $pp$ channels separately, implemented numerically in Refs. \cite{LitvinovaWibowo2018,WibowoLitvinova2019} and \cite{Litvinova2021}, and enabled the astrophysical connection \cite{LitvinovaRobinWibowo2020,Litvinova2021b}. Applications to the electric dipole strength \cite{LitvinovaWibowo2018,WibowoLitvinova2019}, beta decay \cite{LitvinovaRobinWibowo2020}, and electron capture (EC) rates \cite{Litvinova2021b} revealed the necessity of completion of the theory by developing a finite-temperature formalism including superfluidity in a unified way, which is done in Ref. \cite{Bhattacharjee2024}. 

These recent developments determine several directions in which the theory and its computational implementation are to be advanced beyond the state of the art. At the same time, the RNFT is well-positioned to address a variety of nuclear structure phenomena quantitatively, assisting modern high-resolution  
measurements and experiments with exotic nuclei. Some of the recent examples are discussed in this contribution.



\section{Exploring frontiers of nuclear structure phenomenology}  
\subsection{The two-component structure of the pygmy dipole resonance}
\begin{figure}
\centering
\sidecaption
\includegraphics[width=7cm,clip]{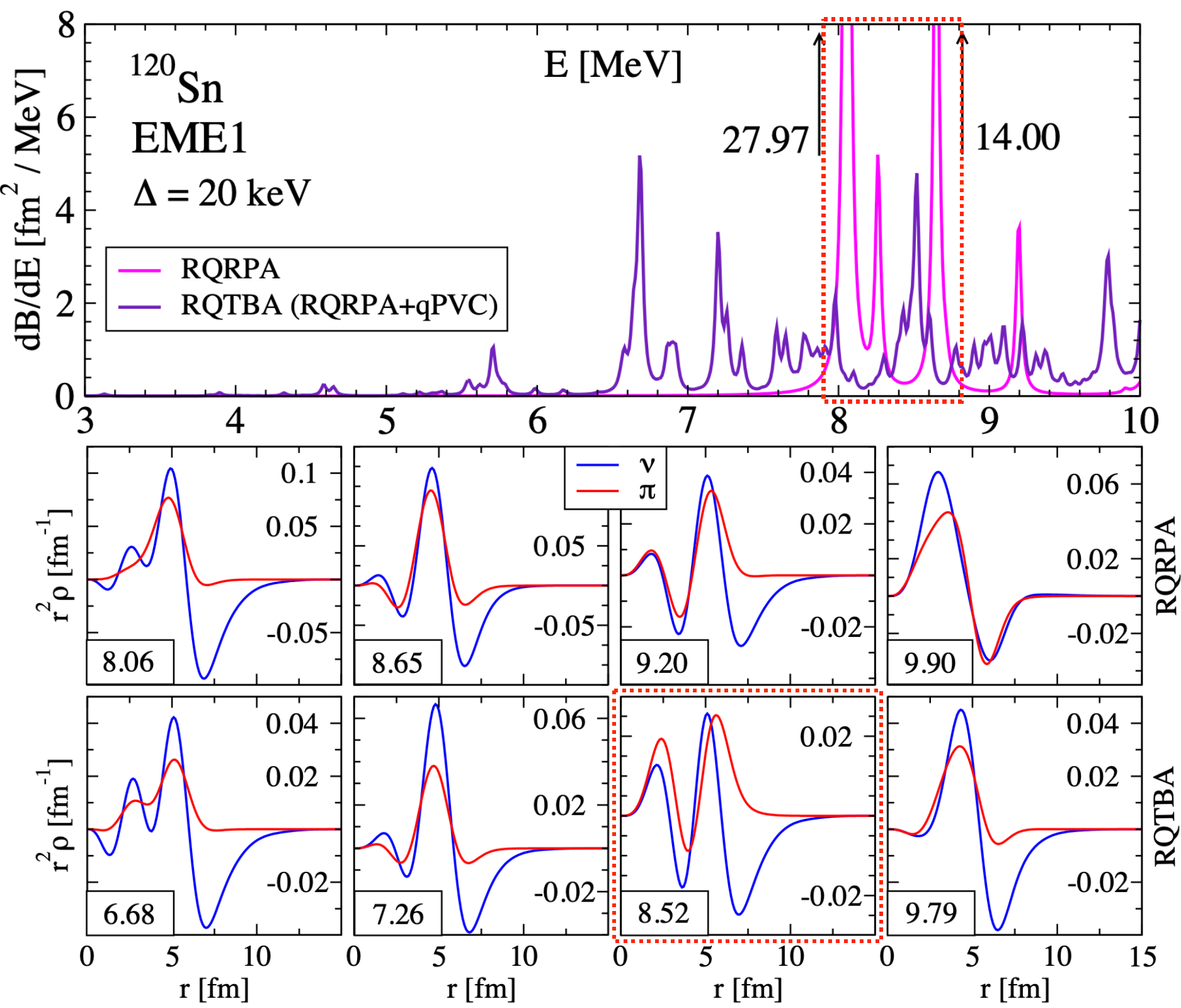}
\caption{ LEDS in 
$^{120}$Sn calculated in RQRPA and RQTBA/REOM$^2$ with $\Delta = $ 20 keV smearing (top panel). The radial neutron ($\nu$) and proton ($\pi$) transition densities $\rho^{(\nu,\pi)}(r)$ extracted from the RQRPA  (middle) and RQTBA (bottom) calculations in the $6 \leq E \leq 10$ MeV energy range. The energies of the states are indicated in the boxes in the bottom-left corners of each panel displaying the transition densities. The dotted box in the top panel encloses the energy interval dominated by the intruder RQTBA states. The figure is reprinted from Ref. \cite{Markova2025}.}
\label{fig-1}       
\end{figure}

The pygmy dipole resonance (PDR) is among the most interesting nuclear vibrational modes because of its relevance for stellar astrophysics 
 \cite{bracco2019isoscalar}. It is a soft, weakly collective excitation on the low-energy tail of the broad high-frequency isovector giant dipole resonance (IVGDR) in the electric dipole (E1) $\gamma$-ray transition probability distribution. While the latter
is associated with the out-of-phase oscillation of the proton and neutron Fermi liquids at 15-20 MeV \cite{harakeh2001giant}, 
the PDR is an E1 strength enhancement grouped around 5-7 MeV in heavy nuclei, bridging with the IVGDR near and below the neutron separation threshold (S$_n$).  Being more pronounced in neutron-rich nuclei, PDR is interpreted as an oscillation of the neutron excess (skin) against a proton-neutron saturated core \cite{mohan1971three,Vretenar2001}, having a mixed isoscalar-isovector character \cite{bracco2019isoscalar, savran2013experimental}, and related to the neutron-skin thickness \cite{piekarewicz2011pygmy}, the symmetry energy in the nuclear matter equation of state \cite{klimkiewicz2007nuclear,Roca2015,fattoyev2018neutron}, and the r-process nucleosynthesis  \cite{goriely1998radiative}. 

\begin{figure}
\centering
\sidecaption
\includegraphics[width=7cm,clip]{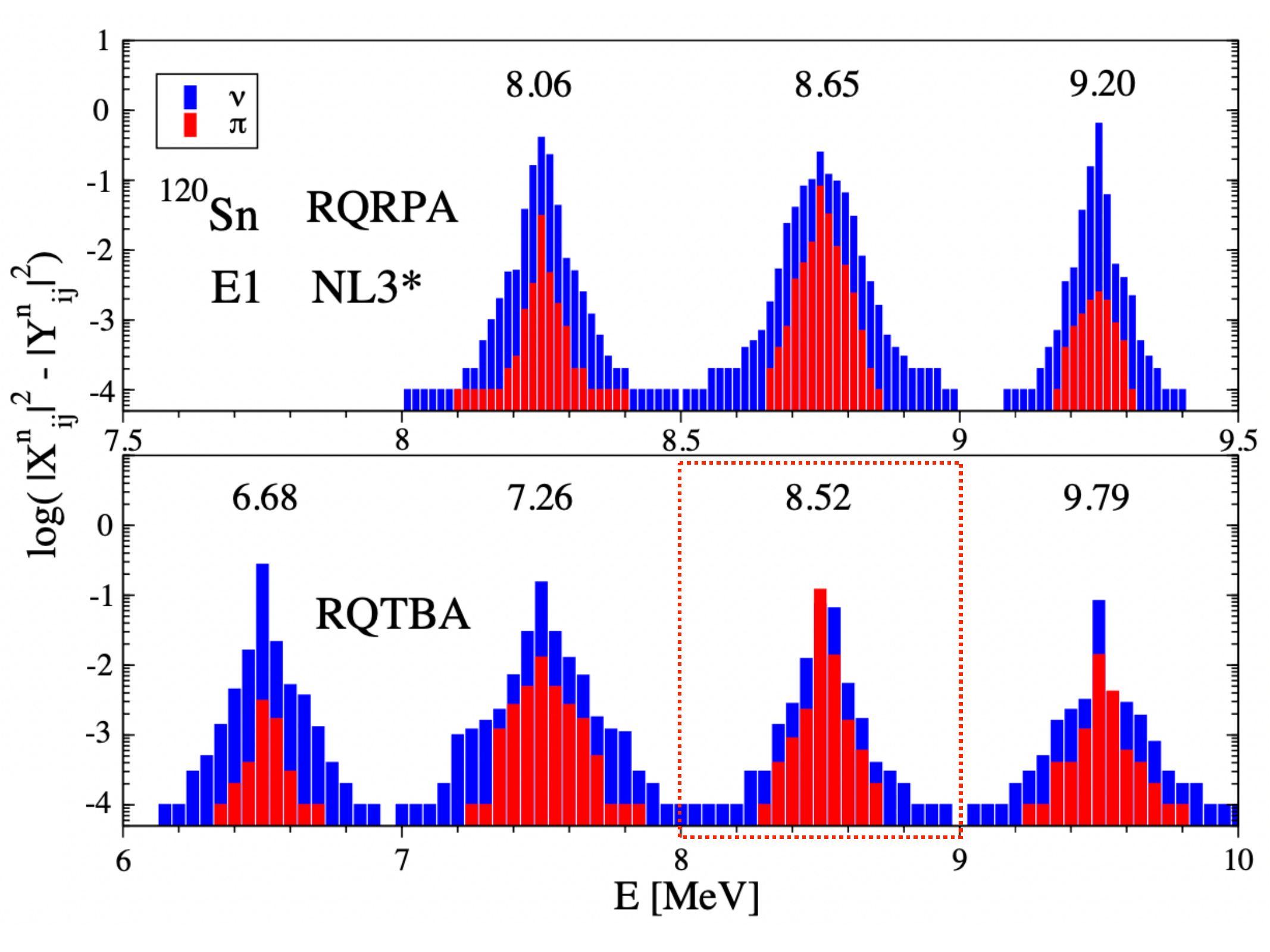}
\caption{The $\cal Z$-values exceeding 0.01\% for the characteristic LEDS states of RQRPA (top) and RQTBA (bottom): ${\cal Z}_{ij} = |{\cal X}^n_{ij}|^2 - |{\cal Y}^n_{ij}|^2$, where $\{i,j\}$ stand for the quasiparticle orbits in the mean-field basis of Dirac-Hartree-Bogoliubov spinors \cite{LitvinovaRingTselyaev2008}. ${\cal X}^{n}_{jk} = \langle 0|\alpha_k\alpha_j|n\rangle$ and ${\cal Y}^{n}_{jk} = \langle 0|\alpha^{\dagger}_{j}\alpha^{\dagger}_k|n\rangle$, where
$\{\alpha^{\dagger}_j,\alpha_j\}$ are the operators of the Bogoliubov quasiparticles, and $|0\rangle$ and $|n\rangle$ are the ground and excited states, respectively. The dotted box encloses the characteristic proton-dominant state. The figure is reprinted from Ref. \cite{Markova2025}.}
\label{fig-2}       
\end{figure}

The fine structure of PDR in even-even $^{112-124}$Sn was investigated systematically in Refs. \cite{Markova2024,Markova2025}. We applied the 
leading-order qPVC, i.e., the relativistic EOM (REOM) limited by $4q$ ($2q\otimes phonon$) configurations, that is, in practice, close to the relativistic quasiparticle time blocking approximation (RQTBA) of Ref. \cite{LitvinovaRingTselyaev2008}, dubbed as RQTBA/REOM$^2$.
This allowed for decomposing the theoretical low-energy dipole strength (LEDS) in $^{116-124}$Sn into two distinct structures, also identified in the experiment.   
As an illustration, the theoretical LEDS for $^{120}$Sn and characteristic transition densities are shown in Fig. \ref{fig-1}.
For each of the 1 MeV energy intervals, a characteristic state with the maximal
$2q$ contribution to the state normalization \cite{LitvinovaRingTselyaev2007,Litvinova2007} was chosen. 
As the leading-order photon field is a single-particle operator, it couples directly to the $2q$ configurations, while the remaining  $2q\otimes phonon$ contributions are manifested through fragmentation of the "primordial" $2q$ configurations of the relativistic QRPA (RQRPA/REOM$^1$). Thus, the RQTBA states with maximal transition amplitudes
correspond to the
most pronounced peaks, which were associated with characteristic states for each 1 MeV energy bin in RQTBA  
and 0.5 MeV in RQRPA. 
The middle and bottom panels of Fig.~\ref{fig-2} display the neutron and proton transition densities $\rho^{(\nu,\pi)}(r)$ of the characteristic states weighted with $r^2$ for RQRPA and RQTBA, respectively. All the RQRPA states below 9.9 MeV are clearly dominated by the neutron skin oscillation, while the $2q$ components of the RQTBA states located between $\sim$7.9 and $\sim$8.8 MeV carry an "intruder" character due to the admixture of the IVGDR mode.  Such states are strongly suppressed in the isoscalar dipole channel, see, for instance, Ref.~\cite{EndresLitvinovaSavranEtAl2010}. 
The $\cal Z$-values quantifying the $2q$ fraction of the characteristic RQTBA states are plotted in Fig. \ref{fig-2}, demonstrating that the intruder states represented by the characteristic state at 8.52 MeV are associated with the dominance of the proton component in their $2q$ composition. 
Further, we considered the leading-order matrix element of the transitions between two excited states  $|m\rangle$ and $|n\rangle$:
 ${\cal F}_{mn} = \langle m|{\cal F}|n\rangle = {f}_{ij}({\cal X}^{m\ast}_{ik}{\cal X}^{n}_{jk} + {\cal Y}^{m\ast}_{jk}{\cal Y}^{n}_{ik})$ (summed over the repeated single-particle indices). The $\gamma$ transition operator $\cal F$ has non-vanishing contributions ${f}_{ij}$ only for the proton components, except for the case of dipole transitions. 
 This indicates that the upper LEDS component, which is dominated by proton $2q$ contributions, has an enhanced probability to gamma decay to a multitude of excited states with higher spin. In contrast, the lower LEDS component related to the neutron skin oscillation predominantly decays to the zero-spin ground state (of an even-even nucleus). The results of Ref. \cite{Markova2025}, thereby, support the idea that only the low-energy part of the LEDS is firmly associated with the neutron skin oscillation mode and suggest revising the previous recommendations on PDR and its impact on the symmetry energy and neutron skin properties. 

\subsection{The nuclear compressibility puzzle}
\begin{figure}
\centering
\sidecaption
\includegraphics[width=7cm,clip]{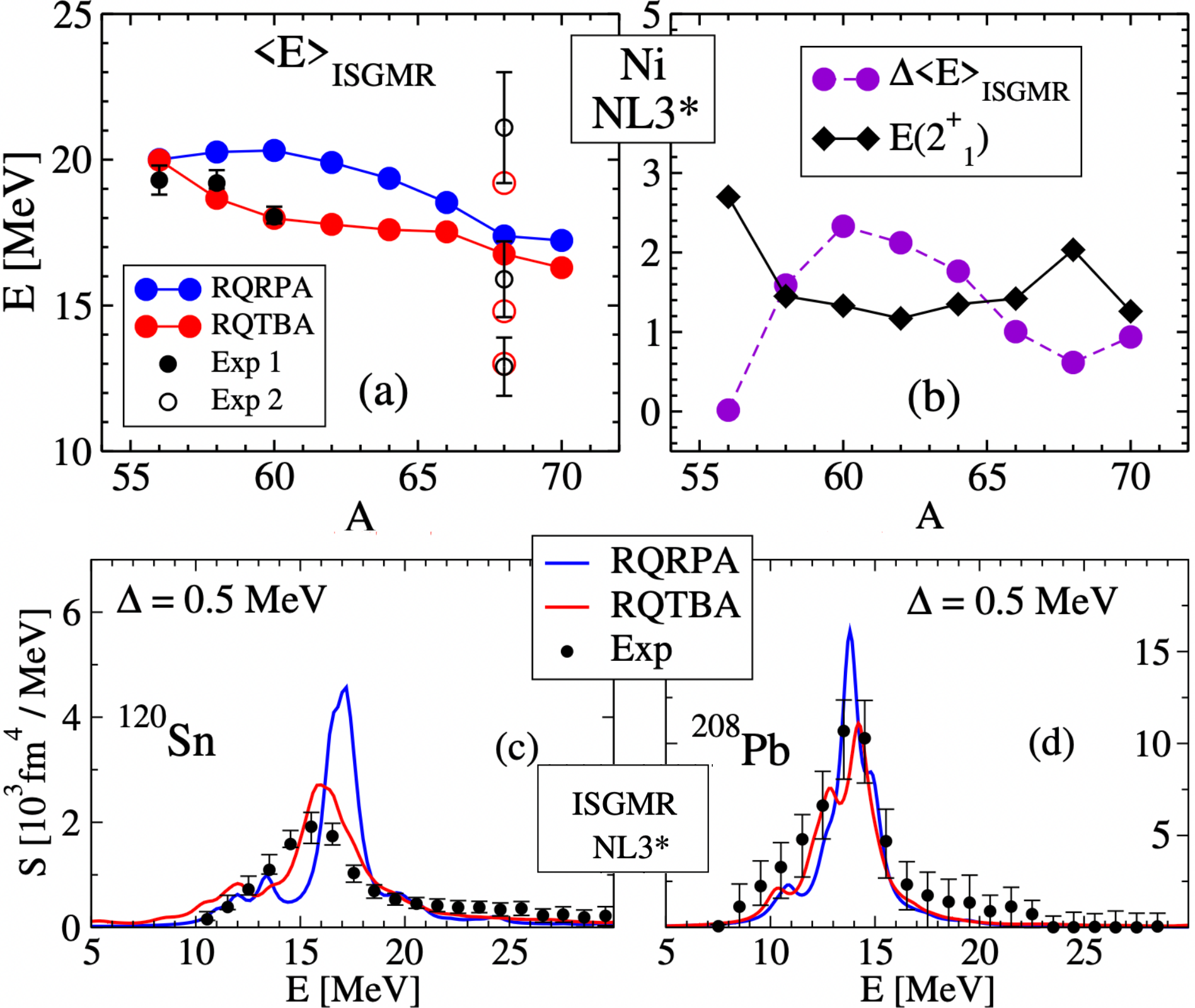}
\caption{(a): ISGMR centroids in nickel isotopes $^{56-70}$Ni compared to data of Refs. \cite{Monrozeau2008} ($^{56}$Ni), \cite{Lui2006} ($^{58,60}$Ni), and \cite{Vandebrouck2014} ($^{68}$Ni).  Empty symbols stand for three separate peaks above 10 MeV determined in \cite{Vandebrouck2014}. (b): the downward shifts $\Delta\langle E \rangle_{\text{ISGMR}}$ of the ISGMR centroids in the RQTBA with respect to the RQRPA ones (circles) in comparison with the E(2$^{+}_1$) values (diamonds). (c,d): The isoscalar monopole response in $^{120}$Sn and $^{208}$Pb as RQRPA and RQTBA strength distributions compared to experimental data \cite{Li2007} ($^{120}$Sn) and \cite{Garg2018} ($^{208}$Pb). The figure is adapted from Ref. \cite{Litvinova2023}.}
\label{fig-3}       
\end{figure}

The nuclear compressibility puzzle was recently addressed in the RNFT framework, applying RQTBA to the monopole response.  To tackle the long-standing problem of the ISGMR description in a single framework throughout the periodic table, a study of the isoscalar giant monopole resonance (ISGMR) in various nuclear systems was conducted in Refs. \cite{Litvinova2023,Li2023}.
The major result of Ref. \cite{Litvinova2023} is that the parameter-free qPVC allows for a realistic description of the ISGMR in nuclei of lead, tin, zirconium, and nickel mass regions, simultaneously. Both the ISGMR widths and centroids were described correctly, which is impossible on the QRPA level. The major puzzle was the ISGMR's centroid, which comes out too high in the so-called soft nuclei, for instance, tin isotopes. The resolution is already possible in the leading qPVC approximation of RQTBA, and Fig. \ref{fig-3} shows selected highlights of this work.

The calculations employed the finite-range effective meson-nucleon interaction NL3* \cite{NL3star}, which, in combination with the qPVC, has consistently demonstrated the ability to reliably describe many other nuclear structure phenomena, including the dipole response discussed above. Systematic calculations of the isoscalar monopole response for nickel isotopes help reveal the central role of the coupling between the ISGMR and the low-energy quadrupole states in the placement of the ISGMR centroids. It appears that, although the placement of the ISGMR centroid generally depends on the nuclear compression modulus $K_{\infty}$ associated with a particular parameter set ($K_{\infty}$ = 258 for NL3*), the coupling of the ISGMR to the low-energy phonons, mostly those of quadrupole character, results in its further fine-tuning. Indeed, one can see a clear inverse correlation between the energies of the lowest quadrupole states $2^+_1$  and the ISGMR centroid shift due to the qPVC effects illustrated in Fig. \ref{fig-3} (a, b).  Panels (c, d) illuminate the contrast between the cases of the soft mid-shell tin nucleus $^{120}$Sn and the doubly-magic $^{208}$Pb associated with weak quadrupole collectivity. Summarizing this case, the coupling between the giant monopole mode and the soft quadrupole mode is largely responsible for spreading and an overall shift of the ISGMR centroid down with respect to its value obtained in the RQRPA. The shifts can amount to 1-2 MeV, they are more pronounced in softer mid-shell nuclei, and the results of the self-consistent RQTBA calculations are in good agreement with data already in the leading qPVC approximation. Further studies with higher configuration complexity approaches are planned for the  ISGMR's fine structure, accessible by high-resolution experiments.

\section{Complex configurations in the nuclear response: toward spectroscopic accuracy}
Quantifying subleading many-body correlations associated with the dynamical kernel of two-fermionic REOMs is of special importance. For decades, (Q)RPA extensions were limited by correlated or uncorrelated $2p2h$  ($4q$) configurations, which include the $2q\otimes phonon$ class. However, even the most rigorous self-consistent implementations of these effects indicate that higher configuration complexity is required for reaching spectroscopic accuracy. Furthermore, including such configurations is needed to clarify deep theoretical aspects of the strongly coupled many-body problem, first of all, the validity of the power counting associated with the emergent scale and calculability of the emergent collective degrees of freedom from the underlying (bare) fermionic interactions. These capabilities may pave the way to transferring the theory across the energy scales and its application to fundamental questions about our universe.

The major conceptual breakthrough was made in Refs. \cite{Litvinova2015,LitvinovaSchuck2019,LitvinovaSchuck2020,Litvinova2022}. In particular, it was shown that the cluster decomposition of the operator strings in the dynamical kernel of the response EOM enables taking into account arbitrarily complex $2nq$ configurations, compatible with factorized forms of the dynamical kernel, by iterative algorithms.
A numerical implementation of this idea to the truncation of the many-body problem at the two-body level, which includes $2q\otimes2phonon$ configurations organized by qPVC, was executed and
 compared to experimental data for medium-light nuclei \cite{LitvinovaSchuck2019,Litvinova2023a,Muescher2024}. It was demonstrated that (i) the $2q$ configurations of RQRPA (identified as REOM$^1$) are able to correctly capture the overall behavior of the IVGDR in terms of its centroid and total strength, while (ii) $2q\otimes phonon$ ($4q$) configurations included in REOM$^2$ are needed for an adequate reproduction of the spreading,  however, the REOM$^2$ strength still shows non-observed gross structures and may underestimate the centroid, so  (iii)  the next level of configuration complexity $2q\otimes 2phonon$ ($6q$) achieved in REOM$^3$ enables eliminating the gross structure artifacts and correcting the centroid, and confidently trends towards the experimental data. Furthermore,  REOM$^n$ with $n > 2$ is necessary to reproduce the observed density of excited states both below and above $S_n$.

Recently, REOM$^3$ was successfully applied to the dipole response of tin isotopes \cite{Novak2024}, where $2q\otimes 2phonon$ configurations enable achieving a smooth distribution of the IVGDR, the feature emerging in heavy nuclei. The method reveals that although $n \sim N$ configuration complexity is formally needed to solve the $N$-body problem exactly, in practice, even at strong coupling, a moderate complexity may be sufficient if the qPVC power counting is used in organizing complex configurations in REOM$^n$. It also verified the robustness of the qPVC power counting across the nuclear landscape, given our preceding REOM$^{1-3}$ calculations for lighter nuclei.
The REOM$^n$ is, thus, an optimal combination of fundamentality, feasibility, and accuracy for the nuclear response with quantified many-body uncertainties. 
Currently, it shows better performance and handling in describing transitional and deformed systems with non-excessive deformations by processing high-rank configurations in a spherical basis, as compared to the finite-amplitude relativistic leading-order qPVC method in the axially-deformed basis completed for the nuclear response following Refs. \cite{Litvinova2022,Zhang2022}. However, these methods are regarded as complementary, and both have the potential to address nuclear spectral properties and push the computational accuracy to a higher degree.  

\section{Nuclear response at finite temperature}

\begin{figure}
\centering
\sidecaption
\includegraphics[width=7cm,clip]{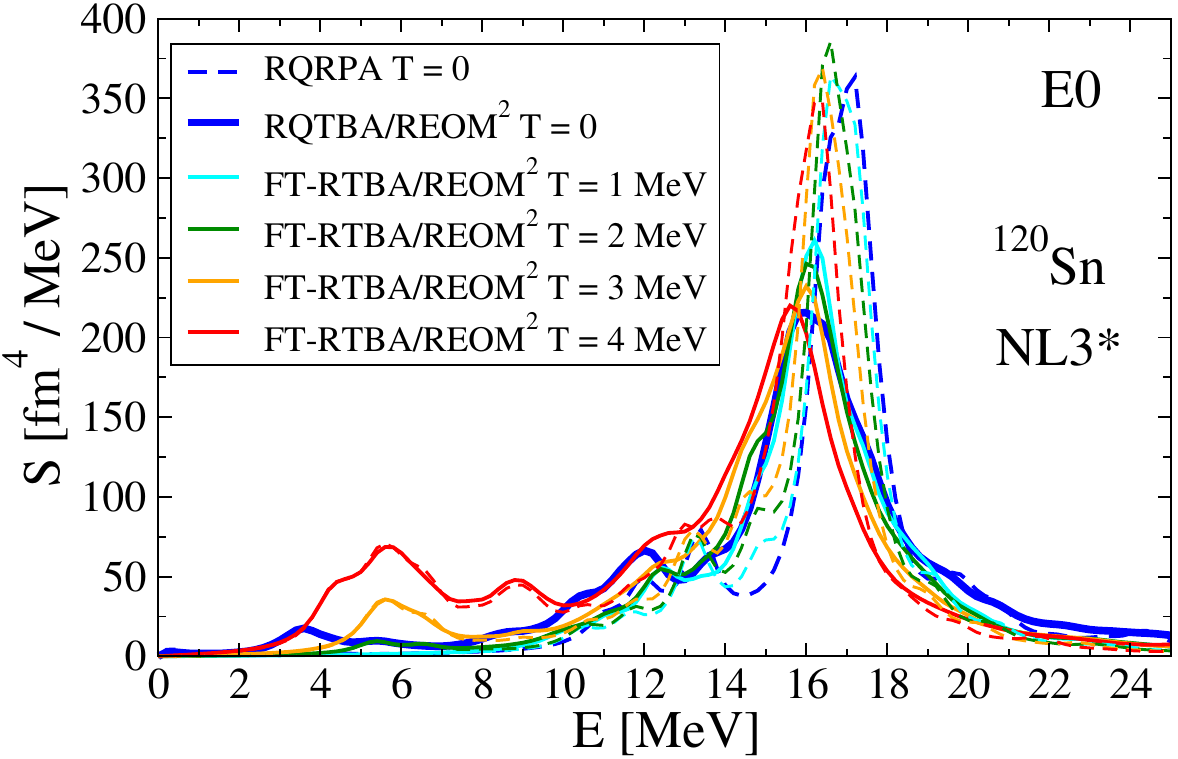}
\caption{Evolution of the monopole response of $^{120}$Sn with temperature. Dashed curves: RQRPA at $T = 0$ and finite-temperature RRPA (FT-RRPA) at $T > 0$. A continuous limit of vanishing superfluidity at the critical temperature $T_c$ in the interval $0 \leq T_c \leq 1$ MeV is assumed following Ref. \cite{Litvinova2021}, for both RQTBA and FT-RTBA. 
The color code of the dashed curves (REOM$^1$) follows that of the solid curves.}
\label{fig-4}       
\end{figure}

The response theory for strongly coupled superfluid fermionic systems at finite temperature was formulated in Ref. \cite{Bhattacharjee2024}. The general many-body Hamiltonian with the bare two-fermion interaction was taken as the starting point, and we conducted a continuous derivation of the EOM for the thermally averaged two-point two-fermion correlation function. The superfluidity was introduced via the most general Bogoliubov transformation of the fermionic field operators; subsequently, the formalism was carried out in the basis of Bogoliubov's quasiparticles, keeping the $4\times 4$ block matrix structure of the EOM. Complete correlated static and dynamical interaction kernels of the resulting EOM were obtained via the commutator technique. The minimally correlated static kernel was linked to the known finite-temperature QRPA. The $4\times 4$ form of the dynamical kernel was previously unknown in both zero-temperature and finite-temperature superfluid theories. 
The major focus was then placed on the latter kernel, which was processed through approximate factorizations keeping fully correlated two-fermion propagators, i.e., truncating the many-body problem at the two-body level. It was advanced explicitly to the generalized qPVC ansatz. In particular, we showed that, as in the zero-temperature and non-superfluid cases, the qPVC emerges with a new order parameter associated with the qPVC vertex, which now acquires an extended form. The latter allows for an efficient organization of complex configurations for building growing complexity converging approximations.

Fig. \ref{fig-4} follows up on the ISGMR in $^{120}$Sn, extending the theoretical study of Ref. \cite{Litvinova2023} to finite temperature. FT-R(Q)RPA  strength distributions $S(E)$ of the isoscalar monopole response are contrasted with FT-R(Q)TBA results at zero and finite temperatures. $S(E)$ is shown before placing the exponential prefactor 1/(1-exp(-$E/T$)) which only modifies the strength below $E\approx T$, where $T$ is the temperature \cite{Sommermann1983,RingRobledoEgidoEtAl1984}. The formation of the low-energy structures at growing temperature was discussed before \cite{Litvinova2018a,LitvinovaWibowo2019}. Here, we are interested in the difference of the ISGMR centroid energies obtained in FT-R(Q)RPA and FT-R(Q)TBA, which was found to be enhanced in this nucleus at $T = 0$. The temperature growth induces the two major effects on the phonons, including the low-energy $2^+$ ones: (i) the transition to the non-superfluid phase, which drastically increases their energies, and (ii) fragmentation leading to the considerable loss of their collectivity \cite{WibowoLitvinova2019}. Accordingly, one can see the visible reduction of the ISGMR centroid shift due to the qPVC when entering the non-superfluid regime. This indicates a sensible temperature dependence of the qPVC's role in nuclear compressibility, which will be studied in a systematic way in future work.

\section{Summary}

Recent progress in the nuclear many-body problem within RNFT was discussed with a special focus on advancements in the configuration complexity of nuclear excited states, superfluidity, and finite temperature. New aspects of the low-energy dipole and high-energy monopole responses were considered and put in the context of previous studies of other nuclear resonant phenomena across the nuclear chart, which are computed   
with the same parametrization of the effective nuclear-meson interaction.  
In the formal sector, a vast amount of subleading correlations is identified for prospective numerical studies. 
Successful benchmarks on available data, appreciable predictive power, and potential for upscaling make RNFT a promising tool for numerous applications at various frontiers of nuclear structure phenomenology and beyond. Furthermore, the universal language and nonperturbative character make the theory transferable across energy scales to study emergent phenomena in other fermionic many-body systems.
Technical problems hindering the achievement of spectroscopic accuracy were described, along with possible solutions, in Ref. \cite{Litvinova2025}.


\section{Acknowledgements}

I thank my collaborators on Ref. \cite{Markova2025} M. Markova and P. von Neumann-Cosel. This work was partly supported by the US-NSF Grants PHY-2209376 and PHY-2515056.

%
\bibliography{BibliographyNov2024.bib} 
%
%
%
%

\end{document}